\documentclass[conference]{IEEEtran}
\IEEEoverridecommandlockouts

\usepackage{amssymb}
\usepackage[cmex10]{amsmath}
\usepackage{stfloats}
\usepackage{graphicx}
\usepackage{subfigure}
\usepackage{tabularx}
\usepackage{epsfig,epsf,color,balance,cite}
\usepackage{verbatim}
\usepackage{url}
\usepackage{bm}
\usepackage{caption}
\usepackage{color}

\newtheorem{theorem}{\bf Theorem}

\newtheorem{proposition}{\bf Proposition}

\usepackage{algorithm}
\usepackage{algorithmic}
\usepackage{color}
\definecolor{myc1}{rgb}{0,0,0}
\begin{document}

% paper title
\title{{Joint Location, Bandwidth and Power Optimization for THz-enabled UAV Communications}  }

\author{
	%\vspace{-50em}
	\IEEEauthorblockN{Luyao Xu, Ming Chen, Mingzhe Chen, Zhaohui Yang, Christina Chaccour, Walid Saad, and Choong Seon Hong
		%,
		%and H. Vincent Poor, \IEEEmembership{Fellow, IEEE}
		\vspace{-4em}
	}
	\thanks{L. Xu  and M. Chen are with National Mobile Communications Research Laboratory, Southeast University, Nanjing, 210096, China, Emails: xuluyao@seu.edu.cn, chenming@seu.edu.cn.}
	\thanks{M. Chen is with the Electrical Engineering Department of Princeton University, NJ, 08544, USA, Email: mingzhec@princeton.edu.}
	\thanks{Z. Yang is with the Department of Engineering, King's College London, WC2R 2LS, UK, Email: yang.zhaohui@kcl.ac.uk.}
	\thanks{C. Chaccour and W. Saad are with the Wireless@VT, Bradley Department of Electrical and Computer Engineering, Virginia Tech, Blacksburg, VA, USA, Emails: christinac@vt.edu, walids@vt.edu.
	}
	\thanks{C. Hong is with the Department of Computer Science and Engineering, Kyung Hee University, Yongin-si, Gyeonggi-do 17104, Rep. of Korea, Email: cshong@khu.ac.kr.}
	%\thanks{H. V. Poor  is  with  the  Department  of  Electrical  Engineering,  Princeton  University,  Princeton,  NJ,  08544,  USA,  Email: poor@princeton.edu.}
}

\maketitle

\begin{abstract}
    In this paper, the problem of unmanned aerial vehicle (UAV) deployment, power allocation, and bandwidth allocation is investigated for a UAV-assisted wireless system operating at terahertz (THz) frequencies. In the studied model, one UAV can service ground users using the THz frequency band. However, the highly uncertain THz channel will introduce new challenges to the UAV location, user power, and bandwidth allocation optimization problems. Therefore, it is necessary to design a novel framework to deploy UAVs in the THz wireless systems. This problem is formally posed as an optimization problem whose goal is to minimize the total delays of the uplink and downlink transmissions between the UAV and the ground users by jointly optimizing the deployment of the UAV, the transmit power and the bandwidth of each user. The communication delay is crucial for emergency communications. To tackle this nonconvex delay minimization problem, an alternating algorithm is proposed while iteratively solving three subproblems: location optimization subproblem, power control subproblem, and bandwidth allocation subproblem. Simulation results show that the proposed algorithm can reduce the transmission delay by up to $59.3\%$, $49.8\%$ and $75.5\%$ respectively compared to baseline algorithms that optimize only UAV location, bandwidth allocation or transmit power control.
\end{abstract}
\IEEEpeerreviewmaketitle

%=========================================   1  ===============================================
\vspace{-.5em}
\section{Introduction}\vspace{0em}
Unmanned aerial vehicles (\,UAVs\,) \cite{8755300} are an effective solution to provide temporary wireless connectivity \cite{8533634,8543138,7756327} due to their flexible maneuverability and convenient deployment. Compared to terrestrial base stations, UAVs can provide higher energy efficiency, improved coverage, and higher wireless capacity. However, deploying UAVs for wireless communications also faces several challenges in terms of deployment, trajectory design, and system optimization.

A number of existing works in \cite{8379427,7918510,7875131,8976230,7486987,9042871,2019Energy} studied the use of UAVs for wireless communications. The authors in \cite{8379427} maximized the number of users that are served by one UAV. In \cite{7918510}, the altitude and location of a UAV were jointly optimized so as to minimize the UAV transmit power used to service ground users. The problem of user-UAV association is studied in \cite{7875131} with the goal of minimizing transmit power. The authors in \cite{8976230} studied the problem of optimal trajectory design and transmit power of the UAV with the goal of maximizing the minimum achievable rate among all users under their harvested energy constraints. In \cite{7486987}, the optimal deployment of multiple UAVs was analyzed to maximize the downlink coverage. Meanwhile, the work in \cite{9042871} studied the problem of three-dimensional (3D) coverage maximization for UAV networks. Moreover, for an UAV enabled mobile edge computing network, the authors in \cite{2019Energy} jointly optimized the location and transmit power of UAVs to minimize the energy consumption. However, none of these existing works such as in \cite{8379427,7918510,7875131,8976230,7486987,9042871,2019Energy} considered the use of the terahertz (THz) band for UAVs. The THz band (0.1-10 THz) is considered to be one of the 6G driving trends owing to its ultra-wide bandwidth \cite{2020A}. Due to the ultra-wide bandwidth of the THz band and the flexible location of the UAV, the combination of UAV and THz can support unprecedentedly high data rate for future services. Nevertheless, given the high uncertainty at higher frequency bands such as THz, it is important to provide more degrees of freedom and control in network management. It is generally known that UAVs can provide line-of-sight (LoS) transmission links to the ground users given their ability to fly. To reap the benefits of UAV deployment in THz systems, it is necessary to optimize the position of the UAVs in a way to provide continual LoS links to the ground users \cite{chaccour2020can}. In \cite{pan2020uav}, the authors proposed the use of a UAV and an intelligent re?ective surface (IRS) to support THz communications while jointly optimizing the UAV’s trajectory, the phase shift of the IRS, the allocation of THz sub-bands, and the transmit power. Different from the work in \cite{pan2020uav}, the authors in \cite{pan2020sum} optimized the location of IRS rather than considering the location of the UAV. However, the works in \cite{pan2020uav} and \cite{pan2020sum} only considered the problem of resource allocation for downlink communication in UAV assisted THz communication, without taking into account the joint downlink and uplink design. Besides, the works in \cite{pan2020uav} and \cite{pan2020sum} only obtained a suboptimal solution for the location of UAV or IRS.

The main contribution of this paper is a first fundamental analysis of the optimal deployment of a UAV that operates at THz frequencies. We consider the downlink and uplink of a wireless network in which THz-enabled UAVs can serve the ground users. Our goal is to minimize the total delays of the uplink and downlink transmissions between the UAV and the users, by jointly optimizing the location of the operating UAV, and the bandwidth of the users, as well as minimizing the transmit power of the users. To solve this optimization problem, we propose an iterative algorithm that divides the original optimization problem into three subproblems: A location optimization subproblem, a power control subproblem, and a bandwith allocation subproblem that are solved iteratively. All the subproblems are shown to be convex, and hence, they can be effectively solved using standard optimization methods. Numerical results show that the proposed algorithm can yield up to $59.3\%$, $49.8\%$ and $75.5\%$ reduction in terms of the transmission delay compared to, respectively, a baseline which only optimizes the UAV location, a baseline which only optimizes the bandwidth of each user, and a baseline which only optimizes the transmit power.

%=========================================   2  ===============================================
\vspace{0em}
\section{System Model and Problem Formulation}
\vspace{0em}%\vspace{-.5em}
% outline
Consider a UAV-assisted wireless communication system that consists of one UAV and $N$ users. The coordinates of the UAV is denoted by $\boldsymbol u_{0}=(x,y,H) $, while the coordinates of each user $ n $ is given by $\boldsymbol u_{n}=(x_{n},y_{n},0) $. Here, the altitude $H$ of the UAV is assumed to be constant.
%\begin{figure}[t]
%    \centering
%    \vspace{-1em}
%    \includegraphics[width=3.5in]{fig1.eps}
%    \vspace{-2.5em}
%    \caption{Architecture of the UAV based THz communication system. % $h_1=9.45\times 10^{-9}$, $h_2=6.17\times 10^{-9}$.
%    } \label{fig1}
%    \vspace{-.25em}
%\end{figure}

The THz channel gain between the UAV and user $ n $ can be modeled as \cite{8763780}:
$h_{n}=d_{n}^{-2} e^{-a\left(f\right) d_{n}}$, where?$d_{n}=\|\boldsymbol u_{n}-\boldsymbol u_{0}\|$?is the distance between user $n$ and the UAV. $e^{-a\left(f\right) d_{n}}$ is the path loss caused by the molecular absorption, and $a\left(f\right)$ is a molecular absorption coefficient that depends on the operating frequency $f$ and concentration of water vapor molecules. For notational simplicity, hereinafter, we use $a$ to represent $a\left(f\right)$. Based on $h_n$, the achievable rate for the uplink transmission from user $n$ to the UAV is \cite{9210812}
\vspace{-.5em}
\begin{equation}\label{sys1equa1}\vspace{-.5em}
r_{n}^{U}={w_{n} \log _{2}\left(1+\frac{p_{n} h_{0}}{w_{n} d_{n}^{2} e^{a d_{n}} \sigma^{2}}\right)},
\end{equation}
where $w_{n}$ is the bandwidth allocated to user $n$, $p_{n}$ is the transmit power of user $n$, $\sigma^{2}$ is the Gaussian noise power, and $h_{0}$?is the channel gain at a reference distance $d_{0}=1$ m.
According to (\ref{sys1equa1}), the uplink transmission delay of user $n$ is\vspace{-.5em}
\begin{equation}\label{sys1equa2}\vspace{-.5em}
t_{n}^{U}\left(p_{n},w_{n},x,y\right)=\frac{D_{n}}{w_{n} \log _{2}\left(1+\frac{p_{n} h_{0}}{w_{n} d_{n}^{2} e^{a d_{n}} \sigma^{2}}\right)},
\end{equation}
where $D_{n}$ is the size of the data that user $n$ needs to transmit to the UAV.
Similarly, the achievable rate for the downlink transmission from the UAV to user $n$ is \vspace{-.5em}
\begin{equation}\label{sys1equa3}\vspace{-.5em}
r_{n}^{D}={w_{n} \log _{2}\left(1+\frac{q h_{0}}{w_{n} d_{n}^{2} e^{a d_{n}} \sigma^{2}}\right)},
\end{equation}
where $q$ is the transmit power of the UAV. Based on (\ref{sys1equa3}), the downlink transmission delay from the UAV to user $n$ is\vspace{-.5em}
\begin{equation}\label{sys1equa4}\vspace{-.5em}
t_{n}^{D}\left(q,w_{n},x,y\right)=\frac{E_{n}}{w_{n} \log _{2}\left(1+\frac{q h_{0}}{w_{n} d_{n}^{2} e^{a d_{n}} \sigma^{2}}\right)},
\end{equation}
where $E_{n}$ is the data that the UAV needs to transmit to user $n$.

Our goal is to minimize the total communication time between the UAV and its served users while satisfying the energy constraints of all users. Mathematically, this optimization problem can be posed as follows:\vspace{-.5em}
\begin{subequations}\label{sys1equa5}\vspace{-.5em}
	\begin{align}
	\min _{\boldsymbol{P},\boldsymbol{W} x, y} &\sum_{n=1}^{N}\left(t_{n}^{U}\left(p_{n},w_{n},x,y\right)+t_{n}^{D}\left(q,w_{n},x,y\right)\right), \tag{\theequation}\\
	\text {s.t.} \quad &\sum_{n=1}^{N} w_{n}=B_W, \label{sys1equa5a}\\
	\quad &p_{n} \leq P, \quad n=1,\cdots,N, \label{sys1equa5b} \\	
	\quad 0 &\leq t_{n}^{U}\left(p_{n},w_{n},x,y\right) p_{n} \leq Q_{n}, \quad n=1,\cdots,N, \label{sys1equa5c}
	\end{align}
\end{subequations}
where $Q_{n}$ is maximum total energy of user $n$, $P$ is maximum transmit power of user $n$, $B_W$ is the total bandwidth of all users, $\boldsymbol{P}=\left[p_{1}, \cdots, p_{N}\right]$, and $\boldsymbol{W}=\left[w_{1}, \cdots, w_{N}\right]$. Constraint (\ref{sys1equa5b}) and (\ref{sys1equa5c}) indicates that each user's energy and transmit power used for data transmission are limited, respectively. Problem (\ref{sys1equa5}) is novel due to the following two reasons. First, the channel model in (\ref{sys1equa1}) takes the molecular absorption into account. Second, the molecular absorption coefficient is a function of distance, which further makes problem (\ref{sys1equa5}) different from previous UAV location optimization problems in \cite{8379427,7918510,7875131,8976230,7486987,9042871,2019Energy} without THz channel.
%=========================================   3  =======================================================================================
\vspace{-1.2em}
\section{Proposed Algorithm}\vspace{-.8em}
% outline
Problem (\ref{sys1equa5}) is a non-convex problem because of the non-convex optimization function and constraints. It is generally challenging to obtain the globally optimal solution for problem (\ref{sys1equa5}). To obtain a suboptimal solution of problem (\ref{sys1equa5}), we propose an algorithm which optimizes $p_{n}$, $w_{n}$, $x$, and $y$ in an alternating manner.
%For the special case with two users, the optimal power control and decoding order is provided in closed form.
\vspace{-1em}
\subsection{User Transmission Power Optimization}\vspace{-.5em}
With fixed $w_{n}$, $x$ and $y$, problem (\ref{sys1equa5}) is formulated as\vspace{-.5em}
\begin{subequations}\label{sys1equa6}\vspace{-.5em}
    \begin{align}
        \min _{\boldsymbol{P}}&\sum_{n=1}^{N}\left(\frac{D_{n}}{w_{n} \log _{2}\left(1+k_{n}(w_{n},x, y) p_{n}\right)}+A_n\right), \tag{\theequation}\\
        \text { s.t. } \quad &\sum_{n=1}^{N} w_{n}=B_W, \label{sys1equa6a}\\
        \quad &p_{n} \leq P, \quad n=1,\cdots,N, \label{sys1equa6b} \\	
        &\frac{\log _{2}\left(1+k_{n}(w_{n},x, y) p_{n}\right)}{p_{n}} \geq l_{n}, \quad n=1,\cdots,N, \label{sys1equa6c}
    \end{align}
\end{subequations}
where $k_{n}(w_{n},x, y)=\frac{h_{0}}{w_{n} d_{n}^{2} e^{a d_{n}} \sigma^{2}}$, $l_{n}=\frac{D_{n}}{w_{n} Q_{n}}$, and $A_{n}=\frac{E_{n}}{w_{n} \log _{2}\left(1+k_{n}(w_{n},x, y) q\right)}$. \vspace{0.2em} It is obvious that constraints and the objective function are convex. Therefore, problem (\ref{sys1equa6}) is convex.

Next, we find the optimal transmit power of each user given the location of the UAV and the bandwidth of each user, as shown in the following proposition.
\begin{proposition}
Given the UAV location $(x,y,H)$ and the bandwidth of each user $w_{n}$, the optimal transmit power of each user $n$ is given by:\vspace{-1em}
\begin{equation}\label{sys1equa7}\vspace{-.8em}
p_{n}^{*}=-\frac{W\left(-\frac{l_{n} \ln 2}{k_{n}(w_{n},x, y)} 2^{-\frac{l_{n}}{k_{n}(w_{n},x, y)}}\right)}{l_{n} \ln 2}-\frac{1}{k_{n}(w_{n},x, y)},
\end{equation}
where function $W\left(x\right)$ is the inverse function of $x e^{x}$.
\end{proposition}
\itshape {Proof:}  \upshape
Since the objective function of problem (\ref{sys1equa6}) is a decreasing function w.r.t. $p_n$, the optimal $p_{n}^{*}$ is the maximum value in the feasible region. According to (\ref{sys1equa6c}), we find that $\frac{\log _{2}\left(1+k_{n}(w_{n},x, y) p_{n}\right)}{p_{n}}$ is a decreasing function w.r.t. $p_n$. Hence, the optimal transmit power $p_{n}^{*}$ satisfies the equation $\frac{\log _{2}\left(1+k_{n}(w_{n},x, y) p_{n}^{*}\right)}{p_{n}^{*}} = l_{n}$. This completes the proof.
\hfill $\Box$

From Proposition 1, we can see that the optimal transmit power of each user $n$ depends on the molecular absorption coefficient, the bandwidth of each user, and the location of the UAV.\vspace{-1em}
\subsection{UAV Location Optimization}\vspace{-.5em}
Given $p_{n}$ and $w_{n}$, the optimization problem in (\ref{sys1equa5}) can be given by:\vspace{-2em}
\begin{subequations}\label{sys1equa8}\vspace{-.5em}
\begin{align}
		\min _{x, y} \sum_{n=1}^{N} &\frac{D_{n}}{w_{n} \log _{2}\left(1+k_{n}(w_{n},x, y) p_{n}\right)} \nonumber \\
		\quad+\sum_{n=1}^{N} &\frac{E_{n}}{w_{n} \log _{2}\left(1+k_{n}(w_{n},x, y) q\right)}, \tag{\theequation}\\
		\text {s.t.} \quad &\sum_{n=1}^{N} w_{n}=B_W, \label{sys1equa8a}\\
		 &p_{n} \leq P, \quad n=1,\cdots,N, \label{sys1equa8b} \\	
        &\frac{D_{n} p_{n}}{w_{n} \log _{2}\left(1+k_{n}(w_{n},x, y) p_{n}\right)} \leq Q_{n}. \label{sys1equa8c}
\end{align}
\end{subequations}
To solve problem (\ref{sys1equa8}), we need to prove that constraint (\ref{sys1equa8c}) and the optimization function in (\ref{sys1equa8}) are convex, we first introduce an important theorem as follows.
\begin{theorem}
	Given $e_{n}$ = $1+k_{n}(w_{n},x, y) p_{n}$ $ < $ 24, function $\frac{D_{n} p_{n}}{w_{n} \log _{2}\left(1+k_{n}(w_{n},x, y) p_{n}\right)}$ is convex w.r.t. $x$ and $y$.
\end{theorem}\vspace{.5em}
\begin{IEEEproof}
	The proof is in the Appendix.
\end{IEEEproof}

Since both the objective function and constraint are convex according to Theorem 1, problem (\ref{sys1equa8}) is convex.

\begin{algorithm}[t]
	\algsetup{linenosize=\small}
	\small
	\caption{Proposed Algorithm for Problem (\ref{sys1equa5})}
	\begin{algorithmic}[1]
		\STATE Initialize $(x^{(1)},y^{(1)},w_{n}^{(1)},p_{n}^{(1)})$, the iteration number $k=1$.
		\REPEAT
		\STATE With given $(x^{(k)},y^{(k)},w_{n}^{(k)})$, obtain $p_{n}^{(k+1)}$ by solving problem (\ref{sys1equa6}).
		\STATE With given $(w_{n}^{(k)},p_{n}^{(k+1)})$, obtain $(x^{(k+1)},y^{(k+1)})$ by solving problem (\ref{sys1equa8}).
		\STATE With given $(x^{(k+1)},y^{(k+1)},p_{n}^{(k+1)})$, obtain $(w_{n}^{(k+1)})$ by solving problem (\ref{sys1equa9}).
		\STATE Set $k=k+1$.
		\UNTIL objective function converges
		\RETURN $x,y,p_{n},w_{n},n=1,2,\cdots,N$
	\end{algorithmic}
\end{algorithm}\vspace{0em}
\subsection{User Bandwidth Optimization}
Given $p_{n}$ and $(x,y,H)$, the optimization problem in (\ref{sys1equa5}) can be given by:\vspace{-.5em}
\begin{subequations}\label{sys1equa9}\vspace{-.5em}
	\begin{align}
		\min _{\boldsymbol{W}} \sum_{n=1}^{N} &\frac{D_{n}}{w_{n} \log _{2}\left(1+k_{n}(w_{n},x, y) p_{n}\right)} \nonumber \\
		\quad+\sum_{n=1}^{N} &\frac{E_{n}}{w_{n} \log _{2}\left(1+k_{n}(w_{n},x, y) q\right)}, \tag{\theequation}\\
		\text {s.t.} \quad &\sum_{n=1}^{N} w_{n}=B_W, \label{sys1equa9a}\\
		&p_{n} \leq P, \quad n=1,\cdots,N, \label{sys1equa9b} \\	
		&\frac{D_{n} p_{n}}{w_{n} \log _{2}\left(1+k_{n}(w_{n},x, y) p_{n}\right)} \leq Q_{n}. \label{sys1equa9c}
	\end{align}
\end{subequations}
It is obvious that constraint (\ref{sys1equa9c}) and the optimization function in (\ref{sys1equa9}) are convex. Therefore, the problem (\ref{sys1equa9}) is convex.\vspace{0em}
\subsection{Algorithm Complexity}\vspace{-.3em}
The overall algorithm used to solve problem (\ref{sys1equa5}) is summarized in Algorithm 1 and Algorithm 1 must converge due to the fact that the objective function with finite lower bound is monotonically nonincreasing. From Algorithm 1, the main complexity of solving problem (\ref{sys1equa5}) lies in solving the power optimization problem in (\ref{sys1equa6}) and the bandwidth allocation problem in (\ref{sys1equa9}). The optimal solution of problem (\ref{sys1equa6}) is obtained in closed-form in (\ref{sys1equa7}), whose complexity is $\mathcal{O}(N)$. Similarly, the complexity of solving problem (\ref{sys1equa9}) also is $\mathcal{O}(N)$. As a result, the total complexity of solving problem (\ref{sys1equa5}) is $\mathcal{O}(2T_{0}{N})$, where $T_{0}$ is the total number of iterations needed to run Algorithm 1.

%=================================   4    ===================================
\vspace{0em}
\section{Simulation Results}\vspace{0em}
For our simulations, a 50 m $\times$ 50 m square area is considered with $N$ = 14 randomly distributed users (unless stated otherwise) and one UAV. Other parameters are listed in Table \uppercase\expandafter{\romannumeral1} \cite{huang2021multi}. All statistical results are averaged over a large number of simulation runs.
\begin{table}[t]
	\centering
	\caption{System Parameters}
\begin{tabular}{ccc}
	\hline
	\hline
	Parameters & Values\\
	\hline
	Gaussian noise power $\sigma^{2}$ &-174 dBm\\
	Channel gain at $d_{0} = 1$ m $h_{0}$ &-40 dB\\
	Molecular absorption coefficient $a$  &0.005 m$^{-1}$\\
	Operating frequency $f$ & 1.2 THz\\
	Data that user $n$ transmits to the UAV $D_{n}$ &[10,8,6,4] Tbits\\
	Data that the UAV transmits to user $n$ $E_{n}$ &[8,6.4,4.8,3.2] Tbits\\
	The transmit power of the UAV $q$ &2 Watts\\
	The total bandwidth of all users $B_W$ &$100$ GHz\\
	Maximum total energy of each user $Q_{n}$ &$Q_{1} = \cdots = Q_{N}=Q$\\
	Maximum transmit power of each user $P$ &0.1 Watts\\
	\hline
	\hline
\end{tabular}
\end{table}

Fig.~1 shows how the transmission delay changes as the molecular absorption coefficient $a$ varies when the total bandwidth of users is $B_W$ = 100 GHz, the maximum total energy of each user $Q$ is 8 J. From Fig.~1, we can see that the transmission delay increases with the molecular absorption coefficient $a$. This is due to the fact that as $a$ increases, the molecular absorption loss increases. From Fig.~1, we can also see that the proposed algorithm can reduce the transmission delay by up to $59.3\%$, $49.8\%$ and $75.5\%$ compared to a first baseline (labeled ‘OL’) that only optimizes the UAV deployment, a second baseline (labeled ‘OP’) that only optimizes the transmit power of the users, and a third baseline (labeled ‘OW’) that only optimizes the bandwidth of each user. Besides, we use the exhaustive search method (labelled 'EXH') to obtain a near global optimal solution, which uses interior point method to jointly optimize transmit power, location and bandwidth with multiple initial solutions. From Fig. 1, the EXH algorithm achieves the best performance at the cost of computation capacity. The gap between the proposed algorithm and the EXH algorithm is small, which indicates that the proposed algorithm can approach the optimal solution.

Fig.~2 shows how the transmission delay changes as the total bandwidth of users vary when the maximum total energy of each user $Q$ is 2 J. From Fig.~2, we can see that the transmission delay decreases as the total bandwidth of users increases. This is because as $B_W$ increases, the achievable rate between the user and UAV increases. We can also see that the transmission delay increases as the UAV's altitude increases under a fixed bandwidth. This is due to the fact that, as the distance between the UAV and users increases, the channel gain decreases.

Fig.~3 shows how the transmission delay changes as the number of users vary when the total bandwidth of users $B_W$ is 100 GHz and the maximum total energy of each user $Q$ is 2 J. From Fig.~3, we can see that the transmission delay increases as the number of users increases and the larger the number of users, the faster the growth rate. This is because as $N$ increases, the distances between the UAV and individual users will be very far.\vspace{-1em}
\begin{figure}[t]
	\centering
	\vspace{-5em}
	\includegraphics[width=3in]{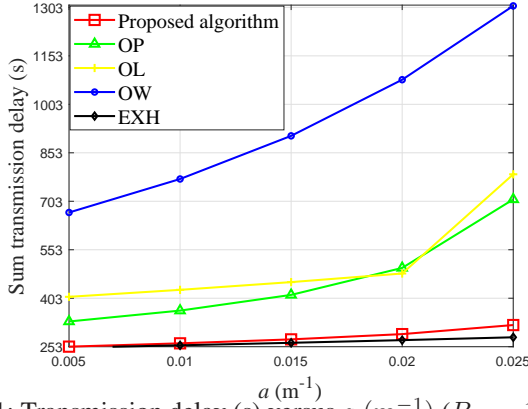}
	\vspace{-.8em}
	\caption*{Fig. 1: Transmission delay (s) versus $a$ $(m^{-1})$ ($B_W$ = 100 GHz, $H$ = 20 m, $Q$ = 8 J).}?
	\label{fig1}
	\vspace{-1.5em}
\end{figure}
\begin{figure}[t]
	\centering
	\vspace{-1em}
	\includegraphics[width=3in]{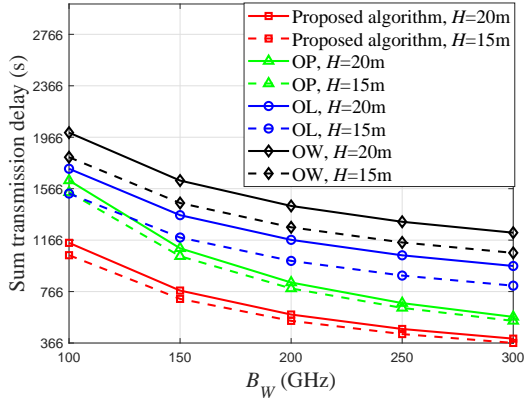}
	\vspace{-.8em}
	\caption*{Fig. 2: Transmission delay (s) versus the bandwidth (GHz) of users  ($Q$ = 2 J).}?
	\label{fig2}
	\vspace{-2.8em}
\end{figure}
\begin{figure}[t]
	\centering
	\vspace{-5em}
	\includegraphics[width=3in]{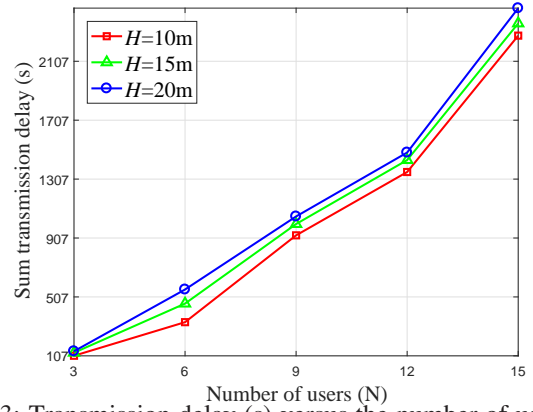}
	\vspace{-.8em}
	\caption*{Fig. 3: Transmission delay (s) versus the number of users ($B_W$ = 100 GHz, $Q$ = 2 J).}?
	\label{fig3}
	\vspace{-2.8em}
\end{figure}
%==============================================    5    ================================================

\section{Conclusion}\vspace{0em}

In this letter, we have proposed a novel approach to deploy a UAV for providing wireless THz connectivity. We have explicitly considered the molecular absorption effect in the THz-enabled UAV channel gain model. Then, we have formulated an optimization problem that seeks to minimize the transmission delay between the UAV and the ground users while meeting the communication requirements of each user. To solve this problem, we have proposed an iterative algorithm that divides the original optimization problem into three subproblems to optimize three variables: the location of the UAV, the users' transmit power and the bandwidth of each user. Simulation results show that the proposed algorithm can significantly reduce the transmission delay compared to three baselines.
\section*{Appendix}\vspace{0em}
We rewrite the function $\frac{D_{n} p_{n}}{w_{n} \log _{2}\left(1+k_{n}(x, y, w_{n}) p_{n}\right)}$ as \vspace{-.5em}
\begin{equation}\label{sys1equa10}\vspace{-.5em}
B\left( e_{n}\right) =\frac{D_{n} p_{n}}{w_{n} \log _{2}(e_{n})}.
\end{equation}
For notational simplicity, hereinafter, we use $B$ to represent $B\left( e_{n}\right)$. The Hessian matrix $\boldsymbol{G}$ of (\ref{sys1equa10}) can be expressed as\vspace{-.5em}
\begin{equation}\label{sys1equa11}\vspace{-.5em}
\boldsymbol{G}=\left[\begin{array}{cc}
\frac{\partial^{2} B}{\partial x^{2}} & \frac{\partial^{2} B}{\partial x \partial y} \\
\frac{\partial^{2} B}{\partial x \partial y} & \frac{\partial^{2} B}{\partial y^{2}}
\end{array}\right].
\end{equation}
Then, to prove that the function $B$ is convex, we need to prove that the Hessian matrix $\boldsymbol{G}$ is positive semi-definite. Hence, we first prove that the first-order determinant of $\boldsymbol{G}$ is positive and then we prove that the second-order determinant of $\boldsymbol{G}$ is positive.

Based on (\ref{sys1equa11}), the first order determinant of $\boldsymbol{G}$ is expressed as follows: \vspace{-.5em}
\begin{equation}\label{sys1equa12}\vspace{-.5em}
\begin{aligned}
\frac{\partial^{2} B}{\partial x^{2}}=&\frac{\partial^{2} B}{\partial d_{n}^{2}}\left(\frac{\partial d_{n}}{\partial x}\right)^{2}+\frac{\partial B}{\partial d_{n}} \frac{\partial^{2} d_{n}}{\partial x^{2}},\\
=&\frac{\partial^{2} B}{\partial e_{n}^{2}}\left(\frac{\partial e_{n}}{\partial d_{n}}\right)^{2}\left(\frac{x-x_{n}}{d_{n}}\right)^{2}\\
+& \frac{\partial B}{\partial e_{n}}\left(\frac{\partial^{2} e_{n}}{\partial d_{n}^{2}}\left(\frac{x-x_{n}}{d_{n}}\right)^{2}+\frac{\partial e_{n}}{\partial d_{n}} \frac{d_{n}^{2}-\left(x-x_{n}\right)^{2}}{d_{n}^{3}}\right),
\end{aligned}
\end{equation}
where $\frac{\partial B}{\partial e_{n}}=-\frac{D_{n} p_{n} \ln 2}{w_{n} e_{n} \ln ^{2} e_{n}}$, $\frac{\partial^{2} B}{\partial e_{n}^{2}}=\frac{D_{n} p_{n} \ln 2\left(\ln e_{n}+2\right)}{w_{n} e_{n}^{2} \ln ^{3} e_{n}}$, $\frac{\partial e_{n}}{\partial d_{n}}=-\frac{h_{0} p_{n}\left(a d_{n}+2\right)}{w_{n} \sigma^{2} d_{n}^{3} e^{a d_{n}}}$, and $\frac{\partial^{2} e_{n}}{\partial d_{n}^{2}}=\frac{h_{0} p_{n}\left(a^{2} d_{n}^{2}+4 a d_{n}+6\right)}{w_{n} \sigma^{2} d_{n}^{4} e^{a d_{n}}}$. Based on (\ref{sys1equa12}), we rewrite $\frac{\partial^{2} B}{\partial x^{2}}$ as \vspace{-.5em}
\begin{equation}\label{sys1equa13}\vspace{-.5em}
\frac{\partial^{2} B}{\partial x^{2}} =I \frac{D_{n} p_{n} \ln 2 \left(e_{n}-1\right)}{w_{n} e_{n}^{2} d_{n}^{4} \ln ^{3} e_{n}},
\end{equation}
where $I=I_{1} z+\left(a d_{n}+2\right) d_{n}^{2} e_{n} \ln e_{n}$, $z=\left(x-x_{n}\right)^{2}$ and $I_{1}=\left(2 e_{n}-\ln e_{n}-2\right)\left(a d_{n}+2\right)^{2}-e_{n} \ln e_{n}\left(a d_{n}+2\right)-2 e_{n} \ln e_{n}$. To prove that $\frac{\partial^{2} B}{\partial X^{2}}$ is positive, we only need to prove that $I>0$.

First, from the expression of $I$, we can treat it as a linear function with slope $I_{1}$. To determine if $I$ is greater than 0, we consider the following two cases:

\emph{Case 1}: If $I_{1}\ge 0$, we can directly obtain $I>0$.

\emph{Case 2}: If $I_{1} < 0$, function $I$ is a decreasing function w.r.t $z$. Since $z = (x-x_{n})^{2} < d_{n}^{2}$, we have $0 \leq z<z_{\max}$, where $z_{\max}=\frac{p_{n} h_{0}}{(e_{n}-1) w_{n} \sigma^2}$. Thus, to prove that $I>0$ in its feasible region, we only need to prove that $z_{\max}<z^{*}$, where $z^{*}$ is the zero point of $I$, and the expression of $z^{*}$ is given by:\vspace{-.5em}
\begin{equation}\label{sys1equa14}\vspace{-.5em}
z^{*}=\frac{\left(a d_{n}+2\right) d_{n}^{2} e_{n} \ln e_{n}}{-I_{1}}.
\end{equation}
From (\ref{sys1equa14}), we find that the minimum value of $z^{*}$ is $z_{\min }^{*}=\frac{\left(a d_{n}+2\right) d_{n}^{2} e_{n} \ln e_{n}}{-I_{1\min}}$, where $I_{1\min}$ is the minimum value of $I_{1}$. If $z_{\min }^{*}>z_{\max }$, then we have $z^{*}>z_{\min }^{*}>z_{\max }$ so that $I>0$. Thus, we first find $I_{1\min}$, then compare $z_{\min }^{*}$ with $z_{\max }$.

From the expression of $I_{1}$, we can see that $I_{1}$ is a function of $\left(a d_{n}+2\right)$. When $e_{n}>1$, we find that $\left(2 e_{n}-\ln e_{n}-2\right)$ is positive. Hence, the function $I_{1}$ has two solutions, i.e.,\vspace{0.3em} $\frac{e_{n} \ln e_{n} \pm \sqrt{(e_{n} \ln e_{n})^2+8e_{n} \ln e_{n}(2e_{n}-\ln e_{n}-2)}}{2\left(2 e_{n}-\ln e_{n}-2\right)}$. $\frac{e_{n} \ln e_{n}}{2\left(2 e_{n}-\ln e_{n}-2\right)}$ is the\vspace{0.3em} minimum point of function $I_{1}$. Fig.~4(a) shows how the value of function $I_{1}$ varies as $\left(a d_{n}+2\right)$ changes. To determine the minimum point of function $I_{1}$ whether exists in the feasible region, we must compare $\frac{e_{n} \ln e_{n}}{2\left(2 e_{n}-\ln e_{n}-2\right)}$ with $2$ by creating a new function $\frac{e_{n} \ln e_{n}}{2\left(2 e_{n}-\ln e_{n}-2\right)}-2$. Fig.~4(b) shows how the value of function $\frac{e_{n} \ln e_{n}}{2\left(2 e_{n}-\ln e_{n}-2\right)}-2$ changes as $e_{n}$ varies. According to the increasing trend and its feasible region, we observe that $\frac{e_{n} \ln e_{n}}{2\left(2 e_{n}-\ln e_{n}-2\right)}$ is less than $2$ when $1<e_{n}<2940.74$. Since the distance between the user and UAV is usually larger than 10 m, $e_{n} < 24$\footnote[1]{For the typical value, $d_{n} \geq$ 10 m, $a$ = 0.005 m$^{-1}$ at $f$ = 1.2 THz, $h_{0}$ = -40 dB, $\sigma^{2}$ = -174 dBm, $w_{n}$ = 10 GHz and $p_{n}$ = 0.001 W, we always have $e_{n} \leq 23.781$.}. As a result, $\frac{e_{n} \ln e_{n}}{2\left(2 e_{n}-\ln e_{n}-2\right)} < 2$. Thus, according to Fig.~4(a), we observe that $I_{1}$ increases as $\left(a d_{n}+2\right)$ increases in its feasible region.
\begin{figure*}[t]
	\centering
	\vspace{-5em}
	\subfigure{
		\begin{minipage}{5cm}
			\centering
			\includegraphics[width=2in]{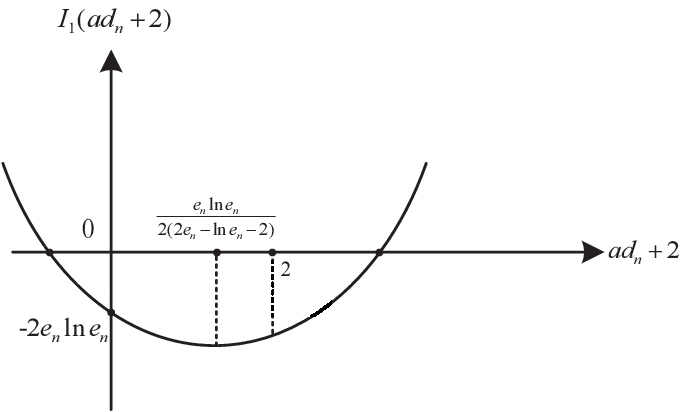}	
			\caption*{(a)}
			\label{fig4(a)}
	\end{minipage}}
	\subfigure{
		\begin{minipage}{5cm}
			\centering
			\includegraphics[width=2in]{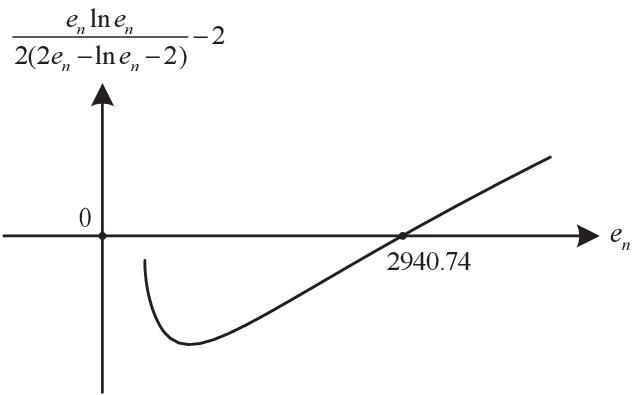}
			\caption*{(b)}
			\label{fig4(b)}	
	\end{minipage}}
	\subfigure{
		\begin{minipage}{5cm}
			\centering
			\includegraphics[width=2in]{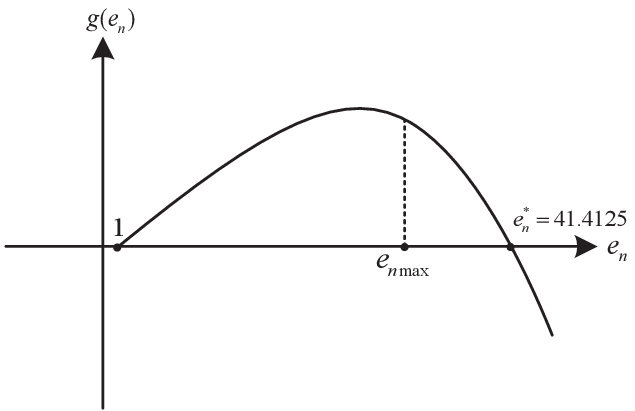}	
			\caption*{(c)}
			\label{fig4(c)}
	\end{minipage}}	
	\vspace{-.5em}
	\caption*{Fig. 4: Illustrations of three functions needed for the analysis of the first order of the determinant.}
	\vspace{-1em}
\end{figure*}
Then, we have:\vspace{-.5em}
\begin{equation}\label{sys1equa15}\vspace{-.5em}
\begin{aligned}
z_{\min }^{*} &=\frac{\left(a d_{n}+2\right) d_{n}^{2} e_{n} \ln e_{n}}{-\left.I_{1}\right|_{\min}} \\
&>\frac{p_{n} h_{0} e_{n} \ln e_{n}}{2 w_{n} \sigma^{2}\left(e_{n}-1\right) \left(\ln e_{n}+2+e_{n} \ln e_{n}-2 e_{n}\right)}.
\end{aligned}
\end{equation}
To compare the value of $z_{\min }^{*}$ with $z_{\max }$, we have:  \vspace{-.5em}
\begin{equation}\label{sys1equa16}\vspace{-.5em}
\begin{aligned}
%=& \frac{h_{0} p_{n} e_{n} \ln e_{n}}{2 w_{n} \sigma^{2}\left(e_{n}-1\right)\left(2+\ln e_{n}+e_{n} \ln e_{n}-2 e_{n}\right)} \\
%&-\frac{h_{0} p_{n}}{w_{n} \sigma^{2}\left(e_{n}-1\right)} \\
z_{\min }^{*}-z_{\max }>& \frac{h_{0} p_{n} g\left(e_{n}\right)}{2 w_{n} \sigma^{2}\left(e_{n}-1\right)\left(2+\ln e_{n}+e_{n} \ln e_{n}-2 e_{n}\right)},
\end{aligned}
\end{equation}
where $g\left(e_{n}\right)=4 e_{n}-4-e_{n} \ln e_{n}-2 \ln e_{n}$. From (\ref{sys1equa16}), we can see that the values of $z_{\min }^{*}$ and $z_{\max }$ depend on the value of $g\left(e_{n}\right)$. Therefore, we next analyze the function $g\left(e_{n}\right)$.

By finding the first and second derivatives of $g\left(e_{n}\right)$, we obtain the trend of $g\left(e_{n}\right)$ in Fig.~4(c). As shown in Fig.~4(c), we define $e_{n}^{*}=41.4125$ as the non-zero point of function $g\left(e_{n}\right)$ and $e_{n\max}$ as the maximum value of $e_{n}$. Since $e_{n}<41$, we have $e_{n}<e_{n}^{*}$. As a result, we have $g\left(e_{n}\right)>0$. Since $g\left(e_{n}\right)>0$, then $z_{\min }^{*}>z_{\max }$ and hence we have $I>0$. Finally, based on (\ref{sys1equa13}), we obtain that the first order determinant of $\boldsymbol{G}$ is positive.

Next, we prove that the second-order determinant of $\boldsymbol{G}$ is positive. The second order determinant of $\boldsymbol{G}$ is expressed as\vspace{-.5em}
\begin{equation}\label{sys1equa17}\vspace{-.5em}
\begin{aligned}
G_{1} &=\left|\begin{array}{cc}
\frac{\partial^{2} B}{\partial x^{2}} & \frac{\partial^{2} B}{\partial x \partial y} \\
\frac{\partial^{2} B}{\partial x \partial y} & \frac{\partial^{2} B}{\partial y^{2}}
\end{array}\right|, \\
%&=\frac{\partial^{2} B}{\partial X^{2}} \frac{\partial^{2} B}{\partial Y^{2}}-\left(\frac{\partial^{2} B}{\partial X \partial Y}\right)^{2}, \\
&=\frac{\left(D_{n} p_{n}\left(e_{n}-1\right) \ln 2\right)^{2}\left(2+a d_{n}\right) L}{w_{n}^{2} e_{n}^{3} \ln ^{5} e_{n} d_{n}^{6}},
\end{aligned}\vspace{.6em}
\end{equation}
where $L=(d_{n}^{2}-H^{2}) S+H^{2} e_{n} \ln e_{n} (2+a d_{n})$ and $S=\left(2 e_{n}-2-\ln e_{n}\right)\left(2+a d_{n}\right)^{2}-2 e_{n} \ln e_{n}$. To prove $G_{1}>0$, we only need to prove that $L$ is positive. Since $S$ is an increasing function w.r.t. $a$, we have $L>\left(d_{n}^{2}-H^{2}\right) S_{1}+H^{2} e_{n} \ln e_{n}\left(2+a d_{n}\right)$ where $S_{1}=S|_{a=0}=2 g\left(e_{n}\right)>0$. As a result, we have  $G_{1}>0$.

Since both the first and second order determinants of $\boldsymbol{G}$ are positive, the Hessian matrix $\boldsymbol{G}$ is positive semi-definite. As a result, function $\frac{D_{n} p_{n}}{w_{n} \log _{2}\left(1+k_{n}(x, y,w_{n}) p_{n}\right)}$ is convex w.r.t. $x,y$. This completes the proof.
\renewcommand\refname{References}
\bibliographystyle{IEEEtran}
\bibliography{IEEEabrv,MMM}

% that's all folks
\end{document}